\documentclass{elsart5p}

 \usepackage{graphicx}
\usepackage{amssymb}

\journal{New Astronomy}

\begin{document}

\begin{frontmatter}



%\title{Spin-powered very high energy emission from the BL Lac \\ 1ES 0806+524}
\title{Is very high energy emission from the BL Lac 1ES 0806+524 \\
centrifugally driven?}


\author{Osmanov Z.} \ead{z.osmanov@astro-ge.org}
% \corauth[cor1]{Corresponding author.}
%\author[Tbilisi1]{D. Lomiashvili}
%\ead{lomiashvili@gmail.com}

\address{E. Kharadze Georgian National Astrophysical Observatory,\\
   Ilia Chavchavadze State University, Kazbegi str. 2a, 0106 Tbilisi, Georgia}

\begin{abstract}
   We investigate the role of centrifugal acceleration of electrons
   in producing the very high energy (VHE) radiation from the BL Lac object
   1ES 0806+524, recently detected by
   VERITAS. The efficiency of the inverse Compton scattering (ICS)
   of the accretion disk thermal
   photons against rotationally accelerated electrons is examined.
   By studying the dynamics of centrifugally induced outflows and by
   taking into account a cooling process due to the ICS, we estimate the maximum
   attainable Lorentz factors of particles and derive corresponding energetic
   characteristics of the emission.
   Examining physically reasonable parameters, by considering
   the narrow interval of inclination angles ($0.7^o-0.95^o$) of magnetic field lines
   with respect to the rotation axis, it is shown that the centrifugally
   accelerated electrons may lead to the observational pattern of the VHE
   emission, if the density of electrons is in a certain interval.
\end{abstract}


\end{frontmatter}

% main text
%%%%%%%%%%%%%%%%%%%%%%%%%%%%%%%%%%%%%%%%%%%%%%%%%%%%%%%%%%%%%%%%%%%%%
\section{Introduction}
%%%%%%%%%%%%%%%%%%%%%%%%%%%%%%%%%%%%%%%%%%%%%%%%%%%%%%%%%%%%%%%%%%%%%

In physics of active galactic nuclei (AGNs) one of the major
problems is related to the understanding of origin of the high
energy radiation. One prominent class of AGNs is the so-called BL
Lac objects - supermassive black holes characterized by rapid and
large amplitude flux variability. By using the radio observations
from the Green Bank 91-m telescope \cite{becker}, the AGN, 1ES
0806+524, was identified as a BL Lac object \cite{schachter}.

%with the photon index $3.6\pm1.0_{stat}\pm 0.3_{sys}$ and the
%following integral flux above $300GeV$ $(2.2\pm0.5_{stat}\pm
%0.4_{sys})\times 10^{12}cm^{-2}s^{-1}$

Recently, by VERITAS was found that the blazar, 1ES 0806+524 reveals
VHE spectra in the TeV domain \cite{veritas}. According to the
standard model of BL Lacs, VHE radiation originates from the ISC of
soft photons against ultra-relativistic electrons
\cite{ktk02,ketal98}. However, the origin of efficient acceleration
of particles up to highly relativistic energies still remains
uncertain and needs to be revealed. Proposed mechanisms based on the
Fermi-type acceleration process \cite{cw99} may be applied
successfully for the TeV emission, only, if the initial Lorentz
factors of electrons are considerably high ($\gamma\geq 10^2$)
\cite{rm00}.

It is clear that in the rotating magnetospheres (the innermost
region of AGN jets and pulsar magnetospheres) the centrifugal effect
should play a significant role in the overall dynamics of
corresponding plasmas. For example, the rotationally driven
parametric plasma instabilities have been studied for pulsars
\cite{incr1,mnras} and AGNs \cite{incr3,drift} respectively, and was
shown, that under certain conditions, the relativistic effects of
rotation may efficiently induce plasma instabilities, parametrically
pumping the rotational energy directly into the plasma waves. The
centrifugally induced outflows have been discussed in a series of
works. Blandford \& Payne in the pioneering paper \cite{bp82}
considered the angular momentum and energy pumping process from the
accretion disk, emphasizing a special role of the centrifugal force
in dynamical processes governing the acceleration of plasmas. It was
shown that the outflows from accretion disks occurred if the
magnetic field lines are inclined at a certain angle to the
equatorial plane of the disk. In the context of studying the
nonthermal radiation from pulsars, the centrifugal effect has been
examined in \cite{g96,g97,tg05}, where the curvature emission of
accelerated particles was studied. By applying the similar approach,
Gangadhara \& Lesch considered the role of centrifugal acceleration
on the energetics of electrons moving along the magnetic field lines
of spinning AGNs \cite{gl97}. This work was reconsidered in a series
of papers \cite{rm00,osm7,ra08} and the method was applied to a
special class of AGNs - TeV AGNs. It was shown that consideration of
straight field lines is a good approximation and was found that the
centrifugal force may accelerate electrons up to very high Lorentz
factors ($\sim 10^8$) providing the TeV energy emission via the ICS.

In the present paper we investigate a role of rotational effects in
the VHE flare from the blazar 1ES 0806+524, by applying the method
of centrifugal outflows, developed in \cite{rm00,osm7,ra08,mr94}. We
show that, for a certain set of parameters, due to the ICS in the
Thomson regime, photons, when upscattered against centrifugally
accelerated ultra-relativistic electrons, produce the VHE radiation
in the TeV domain. We show that a resulting luminosity output is in
a good agreement with the observed data.

The paper is arranged as follows. In \S\ref{sec:con} we consider our
model and derive expressions of the luminosity output and the energy
of photons respectively. In \S\ref{sec:dis} we present the results
for the blazar 1ES 0806+524 and in \S\ref{sec:sum} we summarize our
results.

%%%%%%%%%%%%%%%%%%%%%%%%%%%%%%%%%%%%%%%%%%%%%%%%%%%%%%%%%%%%%%%%%%%%%
\section{Main consideration} \label{sec:con}
%%%%%%%%%%%%%%%%%%%%%%%%%%%%%%%%%%%%%%%%%%%%%%%%%%%%%%%%%%%%%%%%%%%%%

Let us consider the typical parameters of 1ES 0806+524: the black
hole mass, $M_{BH}\approx5\times 10^8M_{\odot}$ \cite{mass},
($M_{\odot}$ is the solar mass) and the bolometric luminosity,
$L\approx 7\times 10^{44}erg/s$ \cite{lumin}. We examine particles
originating from the accretion disc at the distance $\sim 10\times
R_g$ from the central object, where $R_g \equiv 2GM_{BH}/c^2$ is the
gravitational radius of the black hole. Then, by taking the value of
the equipartition magnetic field,
\begin{equation}
\label{b} B\approx \sqrt{\frac{2L}{r^2c}},
\end{equation}
into account, one can show that for typical parameters, $r\approx
10\times R_g$, $n\in (0.0001-1) cm^{-3}$, $\gamma_0\approx 1$, the
value of the ratio, $B^2/\gamma_0mnc^2$, is in the following
interval $\sim 10^{9}-10^{13}$ ($\gamma_0$, $n$ and $m$ are
electrons' initial Lorentz factor, the density and the rest mass
respectively). Therefore, the magnetic field energy density exceeds
the plasma energy density by many orders of magnitude, which
indicates that the plasma co-rotates with the angular velocity,
\begin{equation}
\label{omega} \omega = \sqrt{\frac{GM_{BH}}{r_0^3}},
\end{equation}
corresponding to the Keplerian motion at $r_0\approx 10\times R_g$.

We see that due to the frozen-in condition the particles follow the
co-rotating magnetic field lines and accelerate centrifugally.
Therefore, it is reasonable to consider dynamics of the electron,
sliding along the rotating magnetic field lines. We apply the method
developed for AGNs in \cite{osm7,ra08} and assume that the straight
field lines co-rotate. Then, if we take an angle $\alpha$ between
the magnetic field, $\vec{B}$, and the angular velocity of rotation,
$\vec{\omega}$, into account, after the transformation of
coordinates: $x=rsin\alpha cos\omega t$, $y=rsin\alpha sin\omega t$
and $z=rcos\alpha$ of the Minkowskian metric, $ds^2={\eta}_{\mu
\nu}dx^{\mu}dx^{\beta}$ (${\eta}_{\mu \nu} \equiv diag\{-1,+1,+1,+1
\}$ and $x^{\mu}\equiv (ct;x,y,z)$), the metric in the co-moving
frame of reference is given by \cite{osm7,mr94}
\begin{equation}
\label{metr} ds^2=-c^2\left(1-\frac{\omega^2
r^2sin^2\theta}{c^2}\right)dt^2+dr^2.
\end{equation}

For the equation of motion we get:
\begin{equation}
\label{eq1} \frac{d}{d\chi}\frac{\partial
\mathcal{L}}{\partial\left(\frac{d\bar{x}^{\mu}}{d\chi}\right)}=\frac{\partial
\mathcal{L}}{\partial\bar{x}^{\mu}},
\end{equation}

\begin{equation} \label{lag} \mathcal{L}=-\frac{1}{2}mc\bar{g}_{\mu
\nu}\frac{d\bar{x}^{\mu}}{d\chi}\frac{d\bar{x}^{\nu}}{d\chi},
\end{equation}
\begin{equation} \label{gab}
\bar{g}_{\mu\nu} \equiv diag\left\{-\left(1-\frac{\Omega^2
r^2}{c^2}\right),1\right\},
\end{equation}
where
$$ \Omega = \omega\sin\alpha,\;\bar{x}^{\mu}\equiv (ct;r).
$$

Then, by taking the four velocity identity,
$\bar{g}_{\alpha\beta}\frac{d\bar{x}^{\alpha}}{d\chi}
\frac{d\bar{x}^{\beta}}{d\chi}=-1$, into account, one can derive
from Eq. (\ref{eq1}) the radial equation of motion \cite{mr94}:
\begin{equation}
\label{eul_0}
\frac{d^2r}{dt^2}=\frac{\Omega^2r}{1-\frac{\Omega^2r^2}{c^2}}\left[1-\frac{\Omega^2r^2}{c^2}
-\frac{2}{c^2}\left(\frac{dr}{dt}\right)^2\right].
\end{equation}

Solving Eq. (\ref{eul_0}), it is straightforward to show that the
Lorentz factor of the particle changes radially as \cite{rm00}:
\begin{equation}\label{gama}
\gamma =  \frac{1}{ \sqrt{\widetilde{m}}
\left(1-\frac{r^2}{R_{lc}^2}\right)},
\end{equation}
where
$$\widetilde{m} =\frac{1-r_0^2/R_{lc}^2-\upsilon_0^2/c^2}{\left(
1-r_0^2/R_{lc}^2\right)^2},\; R_{lc}\equiv \frac{c}{\omega}.$$
$r_0$ and $\upsilon_0$ are the initial position and the initial
radial velocity of the particle, respectively and $R_{lc}$ is the
radius of the light cylinder - a hypothetical zone, where the linear
velocity of rigid rotation exactly equals the speed of light, $c$.

As is clear from Eq. (\ref{gama}), in due course of time the Lorentz
factors of electrons become very high in the vicinity of the light
cylinder ($r\sim R_{lc}$). On the other hand, it is clear that
acceleration lasts until the electron encounters a photon, which in
turn inevitably limits the Lorentz factor of the particle. During
the ICS an electron will lose energy, whereas a photon will gain
energy. This mechanism is characterized by the so-called cooling
timescale \cite{rybicki}
\begin{equation}\label{cool}
t_{cool}= 3\times\frac{\gamma}{\left(\gamma^2 - 1\right)U_{rad}}[s]
\,,
\end{equation}
where $U_{rad} = L/4\pi cr^2$ is the energy density of the
radiation. The acceleration process is characterized by the
acceleration timescale, $t_{acc}\equiv\gamma/(d\gamma/dt)$, which
after applying Eq. (\ref{gama}) can be presented by
\begin{equation}\label{tacc}
t_{acc} =
\frac{c\sqrt{1-\frac{\Omega^2r^2}{c^2}}}{2\Omega^2r\sqrt{1-\widetilde{m}
\left(1-\frac{\Omega^2r^2}{c^2}\right)}} \,.
\end{equation}
Generally speaking, initially the electrons accelerate, but in due
course of time the role of the inverse Compton losses increase and
the acceleration becomes less efficient. The maximum energy
attainable by electrons is achieved at a moment when the energy gain
is balanced by the energy losses due to ICS. Mathematically this
means that the following condition $t_{acc}\approx t_{cool}$ has to
be satisfied. After applying Eqs. (\ref{cool},\ref{tacc}) the
aforementioned condition leads to the expression of the maximum
Lorentz factor \cite{rm00}
\begin{equation}\label{gmax}
\gamma_{max}\approx
10^{14}\sqrt{\widetilde{m}}\left[\frac{6\Omega}{U_{rad}(R_{L})}\right]^2
\,,
\end{equation}
where $R_{l}\approx R_{lc}/\sin\alpha$. If electrons with such high
kinetic energies encounter soft photons having energy, $\epsilon_s$,
then, photons' energy after scattering is given by
\begin{equation}\label{ener}
\epsilon = \gamma_{max}^2\epsilon_s \,.
\end{equation}
As we have already mentioned, the particles reach maximum kinetic
energy almost on the LC surface. Let us assume that a layer where
the ICS takes place and the high energy photons are produced has a
thickness, $\Delta r$. Then, for the corresponding infinitesimal
volume of a cylindrical layer we get:
\begin{equation}\label{dv}
dV \approx \frac{\pi R_{lc}(2R_{lc}+\Delta r)\Delta
r}{\sin\alpha}d\alpha \,.
\end{equation}
If we take a single particle Thomson power
\begin{equation}\label{pow}
P_{IC} \simeq \sigma_Tc\gamma^2U_{rad} \,,
\end{equation}
into account, then the total power emitted from the radiation zone
can be expressed as follows
$$L_{IC} = \int
nP_{IC}(\alpha)dV=$$
\begin{equation}\label{power}
=\sigma_T\pi cn R_{lc}(2R_{lc}+\Delta r)\Delta
r\int_{\alpha_1}^{\alpha_2}\frac{\gamma^2_{max}(\alpha)
U_{rad}(\alpha)}{\sin\alpha} d\alpha \,,
\end{equation}
where $\sigma_T\approx 6.65\times 10^{-25}cm^{2}$ is the Thomson
cross-section.

%%%%%%%%%%%%%%%%%%%%%%%%%%%%%%%%%%%%%%%%%%%%%%%%%%%%%%%%%%%%%%%%%%%%%
\section{Discussion} \label{sec:dis}
%%%%%%%%%%%%%%%%%%%%%%%%%%%%%%%%%%%%%%%%%%%%%%%%%%%%%%%%%%%%%%%%%%%%%

According to the observations of VERITAS, performed from November
2006 to April 2008, the blazar 1ES0806+524, has the VHE emission
with the intrinsic photon spectrum, $\Gamma_{i} = (2.8\pm 0.5)$, and
the intrinsic integral flux, $\mathcal{F}_{_{>0.3TeV}} =
(4.4\pm1.1)\times 10^{-12}cm^{-2}s^{-1}$ \cite{veritas}. Then the
intrinsic differential photon spectrum can be easily restored by the
integral flux

$$\frac{dN_{i}}{d\epsilon} =
\left(\frac{3}{4}\right)^{\Gamma_{i}}\times\frac{\Gamma_{i}-1}{0.3TeV}\times\mathcal{F}_{_{>0.3TeV}}\times\left(\frac{\epsilon}{0.4TeV}\right)^{-\Gamma}
=$$
\begin{equation}\label{spectr}
= F_0\times \left(\frac{\epsilon}{0.4TeV}\right)^{-\Gamma_i} \,,
\end{equation}
where $F_0\approx(11.8\pm2.9)\times 10^{-12}cm^{-2}s^{-1}$.

The corresponding luminosity above $0.3TeV$ up to $1TeV$ can be
estimated as follows:
\begin{equation}\label{lum1}
L_{_{VHE}}=\Delta
S\int_{\epsilon_1}^{\epsilon_2}\frac{dN_i}{d\epsilon}\epsilon
d\epsilon \,,
\end{equation}
where $\epsilon_1 = 0.3TeV$ and $\epsilon_2 = 1TeV$. We assume that
the high energy emission originates from the jet, having an opening
angle, $2\alpha_{m}$ (where $\alpha_m\geq\alpha_2$). $\Delta
S\approx \pi D^2\left(\sin^2\alpha_2-\sin^2\alpha_1\right)$ and
$D\approx 630Mpc$ is the distance to the blazar.

By taking the parameters into account, one can see from Eq.
(\ref{lum1}) that the luminosity in the energy interval $(0.3-1)TeV$
is given by
\begin{equation}\label{lum}
L_{_{VHE}}\approx (3.7\pm 1.3)\times
10^{43}\left(\sin^2\alpha_2-\sin^2\alpha_1\right)\frac{erg}{s} \,.
\end{equation}

According to the standard theory, it is well known that the
accretion disks thermally radiate and the corresponding temperature
is expressed as in the following way \cite{sak}:
\begin{equation}\label{tem}
T \approx 3.1\times
10^8Q^{2/5}M_8^{-1/5}d^{-3/5}\left(1-d^{-1/2}\right)^{2/5}K\,,
\end{equation}
where
\begin{equation}\label{Q}
Q\equiv\frac{\dot{M}\;yr^{-1}}{3M_{8}\;yr^{-1}} \,,
\end{equation}
is the dimensionless mass accretion rate, $M_8\equiv
M_{BH}/10^8M_{\odot}$ and $d\equiv r_*/3R_g$.

For the given luminosity, the mass accretion rate can be estimated
as:
\begin{equation}\label{dm}
\dot{M} \approx \frac{L}{0.1c^2} \,,
\end{equation}
then, combining Eqs. (\ref{tem},\ref{dm}) one can show that energy,
$\epsilon_{s} = kT$, of accretion disc's thermal photons emitted in
the area from $r^* = 15\times R_g$ to $r^* = R_{lc}$ is of order
$\sim 10eV$. Therefore, as we see from Eq. (\ref{ener}), for
producing energies from thousands of $GeV$ to $TeV$ domain, one
requires very high Lorentz factors $(1-3)\times 10^5$.

\begin{figure}
  \resizebox{\hsize}{!}{\includegraphics[angle=0]{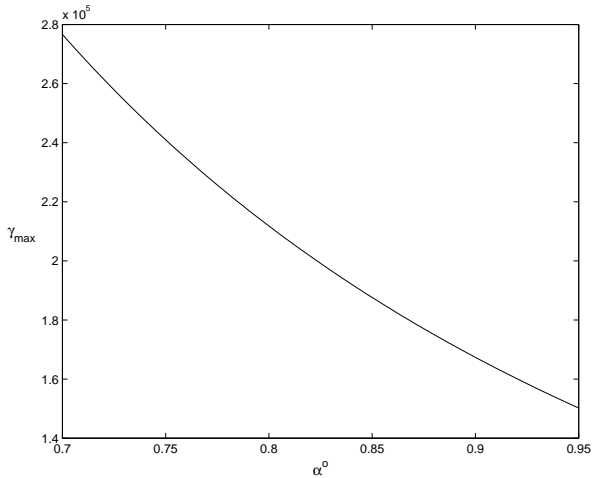}}
  \caption{The behaviour of the maximum Lorentz factors of electrons versus
  the inclination angle.
  The set of parameters is $\upsilon_0 = 0.4c$, $r_0 = 10\times R_g$,
  $\omega = 4.5\times 10^{-6}s^{-1}$ and $L = 7\times 10^{44}erg/s$.}\label{gamma}
\end{figure}

One can see that the aforementioned values of Lorentz factors are
achieved for very low inclination angles.  In Fig. \ref{gamma} we
show $\gamma_{max}$ as a function of the inclination angle. The set
of parameters is $\upsilon_0 = 0.4c$, $r_0 = 10\times R_g$, $\omega
= 4.5\times 10^{-6}s^{-1}$ and $L = 7\times 10^{44}erg/s$. As is
clear from the figure, the electrons reach high values of the
Lorentz factor for small angles, $0.7^o-0.95^o$. This is a natural
result because, one can straightforwardly show from Eq.
(\ref{gmax}), that $\gamma_{max}(\alpha)$ behaves as
$1/\sin^2\alpha$ and therefore, provides higher kinetic energies for
lower inclinations.

The present model is based on an assumption that maximum kinetic
energy of particles is determined by the balance of energy gain due
to the acceleration and energy losses due to the ICS. Generally
speaking, this approach is valid only if the energy losses is
dominated by the ICS. On the other hand, apart from the inverse
Compton scattering, also the curvature radiation could impose
significant limitations \cite{levins}.

The centrifugal acceleration mainly happens close to the light
cylinder, and since the power of a single particle curvature
radiation behaves as $\sim\gamma^4$, one has to check the constraint
imposed by this mechanism on relativistic particle dynamics. A total
power radiated by a single particle is given by
\begin{equation}\label{pcr}
P_c = \frac{2}{3} \frac{e^2 c}{R_c^2} \gamma^4\,,
\end{equation}
where by $R_c$ we denote the curvature radius. Then, the timescale
of curvature emission can be defined by the following way:
\begin{equation}\label{tcr}
t_c = \frac{\gamma mc^2}{P_c}\,.
\end{equation}
To find the limitation imposed on the maximum Lorentz factor let us
note that  electrons initially accelerate efficiently, and this
process lasts until the energy gain is balanced by the curvature
losses. This happens when $t_{\rm acc}\approx t_c$. By taking Eqs.
(\ref{tacc},\ref{tcr}) into account and assuming $R_c\sim R_{lc}$,
it is straightforward to show
\begin{equation}\label{gcr}
\gamma_{max}^{c}\simeq\left(\frac{3mc^3}{e^2\omega\gamma_0^{1/2}}\right)^{2/5}\approx
\frac{3.5\times 10^{11}}{\gamma_0^{1/5}}.\end{equation}

From Eqs. (\ref{gmax},\ref{gcr}) we see that for $\gamma_0\sim 1$
one has the following inequality $\gamma_{max}^{c}\gg\gamma_{max}$.
This indicates that the curvature radiation does not impose a
significant limitation on the maximum attainable Lorentz factors.
Therefore we conclude that maximum attainable kinetic energies are
determined only by the ICS.

As is clear from Eq. (\ref{ener}), the VHE emission ($0.3TeV-1TeV$)
detected by VERITAS can be achieved for the following interval of
Lorentz factors $\left(1.5-2.8\right)\times 10^5$. Equation
(\ref{gmax}) shows that when the inclination angles are in the
following range $\alpha\in(0.7^o-0.95^o)$, electrons attain Lorentz
factors from the aforementioned region of values (see Fig.
\ref{gamma}). $\upsilon_0 = 0.6c$, $r_0 = 10\times R_g$, $\omega =
4.5\times 10^{-6}s^{-1}$ and $L = 7\times 10^{44}erg/s$.
\begin{figure}
  \resizebox{\hsize}{!}{\includegraphics[angle=0]{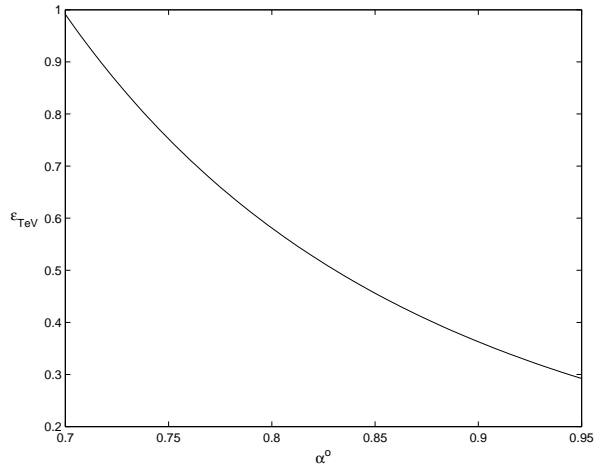}}
  \caption{The dependence of the energy of photons on the inclination angle.
  The set of parameters is $\upsilon_0 = 0.4c$, $r_0 = 10\times R_g$,
  $\omega = 4.5\times 10^{-6}s^{-1}$ and $L = 7\times 10^{44}erg/s$.}\label{energy}
\end{figure}

In Fig. \ref{energy} we show the behavior of the emission energy
versus the inclination angle. As is clear from the figure,
acceleration of electrons inside the region, $0.7^o\leq\alpha^o\leq
0.95^o$, provides the photon energies from $0.3Tev$ up to $1TeV$.

For finding the luminosity output above $0.3TeV$, we have to
integrate Eq. (\ref{power}) from $\alpha_1 \approx 0.7^o$ to
$\alpha_2 \approx 0.95^o$. Let us consider the following set of
parameters: $n = 1.1\times 10^{-3}cm^{-3}$, $\upsilon_0 = 0.4c$,
$r_0 = 10\times R_g$ and $r^* = 15\times R_g$, then, the integration
in Eq. (\ref{power}) gets the VHE luminosity output of order
$4.8\times 10^{39}erg/s$. From Eq. (\ref{lum}) we see that for
$\alpha_1\leq\alpha\leq\alpha_2$, the observed VHE luminosity is in
the range $\sim(4.6\pm 1.6)\times 10^{39}erg/s$, therefore, our
results are in a good agreement with the observed data.

For plotting our graphs and getting the results we used the
parameters with the best fitting, although the values from the
following ranges are also applicable: $r_0/R_g\in [10-20]$,
$\upsilon_0\in [0-0.7c]$, $n\in [0.7-1.4]\times 10^{-3}cm^{-3}$,
$r^*/R_g\in [10-20]$.

On the other hand, the high energy photons may undergo the
$\gamma\gamma$ absorbtion. It is well known that gamma rays interact
most effectively with the background photons of energy \cite{chiang}
\begin{equation}\label{eb}
\epsilon_b = 4\frac{(mc^2)^2}{\epsilon}\approx 1\frac{
TeV}{\epsilon}eV\,
\end{equation}
and the corresponding cross section has a peak at
$\sigma_0\approx\sigma_T/5$. The optical depth of high energy
photons, then becomes
\begin{equation}\label{tau1}
\tau\equiv\frac{R_{lc}}{\lambda} = n_{IR}\sigma_0 R_{lc}\,,
\end{equation}
where $\lambda$ is the mean-free path of infrared photons,
\begin{equation}\label{n}
n_{IR} = \frac{L(\epsilon_b)}{4\pi R_{lc}^2c\epsilon_{b}}\,
\end{equation}
is the corresponding photon density and $L(\epsilon_b)$ - the
infrared luminosity. After substituting Eq. (\ref{n}) into Eq.
(\ref{tau1}), one can derive an expression of the optical depth of
high energy photons \cite{ahar}
\begin{equation}\label{tau}
\tau = \frac{L(\epsilon_b)\sigma_0}{4\pi R_{lc}c\epsilon_b}\approx
5\times\frac{L(1\frac{TeV}{\epsilon}eV)}{10^{-6}L_{Edd}}\times\frac{10^5}{\gamma}
\times\frac{\epsilon}{TeV}\,,
\end{equation}
where $L_{Edd}\approx 6.5\times 10^{46}erg/s$ is the Eddington
luminosity of 1ES 0806+524.

As we see from the figures, for producing radiation in the $1TeV$
domain, one has to accelerate the electrons up to $\gamma\approx
2.8\times 10^5$. For 1ES 0806+524 no infrared data are published so
far, on the other hand, since the TeV emission is detected,
therefore the infrared luminosity of 1ES 0806+524 must be less than
$5.6\times 10^{-7} L_{Edd}\approx 3.6\times 10^{40}erg/s$ (see Eq.
(\ref{tau})). One has to note that the centrifugal acceleration
leads to the TeV variability timescale of the order $\sim (1-2)days$
\cite{ra08}. It is worth noting that TeV blazars exhibit the
variability on hour to minute timescales, but this particular
feature is not detected for 1ES 0806+524.

As our investigation shows, the centrifugally accelerated outflows
may provide the detected VHE emission of 1ES 0806+524 via the ICS if
the parameters are chosen appropriately. From the aforementioned set
of parameters very important one is the density of relativistic
electrons, which has to be in the following interval
$[0.7-1.4]\times 10^{-3}cm^{-3}$ in order for the centrifugal
acceleration to explain the detected high energy emission. This we
consider as a certain test to check if the mentioned mechanism is
feasible.

%%%%%%%%%%%%%%%%%%%%%%%%%%%%%%%%%%%%%%%%%%%%%%%%%%%%%%%%%%%%%%%%%%%%%
\section{Summary} \label{sec:sum}
%
%
%%%%%%%%%%%%%%%%%%%%%%%%%%%%%%%%%%%%%%%%%%%%%%%%%%%%%%%%%%%%%%%%%%%%%

\begin{enumerate}
      \item For explaining the observed TeV energy radiation from
      1ES 0806+624 detected by VERITAS, we have considered the
      inverse Compton scattering of disk thermal photons against
      centrifugally accelerated ultra-high energy electrons.

      \item We have shown that due to very strong magnetic field,
      the electrons are in the frozen-in condition, which leads to
      the co-rotation of particles. Due to the co-rotation, electrons centrifugally
      accelerate almost up to the light cylinder surface, and in the
      nearby zone of it the electrons upscatter against soft
      thermal photons, causing
      the limitation of particles' kinetic energy. We also have shown
      that the role of the curvature radiation in limiting the maximum kinetic
      energy is negligible with respect to the inverse Compton losses.

      \item We considered the $\gamma\gamma$ absorbtion of high energy
      photons in the background field of soft infrared photons.
      From the observationally evident
      fact that the TeV radiation escapes the central source, we estimated the
      maximum value of infrared luminosity which has not been detected so far.

      \item We have found that for physically reasonable parameters
      the ICS occurs in the Thomson regime. It has been
      shown that for the following interval of inclination angles,
      $0.7^o-0.95^o$, with the best fitting
      parameters the resulting emission energies, ($0.3TeV-1TeV$) and
      the luminosity output, $10^{39}erg/s$
      are in a good agreement with the observed data.
      On the other hand, if the parameters are chosen in certain
      physically reasonable intervals, the high energy emission is
      also possible. Therefore we offer the following test:
      if one indirectly measures the density of the relativistic
      electrons, and finds its value
      to be in the following range, $[0.7-1.4]\times 10^{-3}cm^{-3}$, then the centrifugal acceleration is a feasible mechanism
      in producing the TeV photons via the ICS.

\end{enumerate}

In the paper we have done several approximations. The first
limitation is that we studied the straight magnetic field lines,
although, especially in the very vicinity of the light cylinder the
curvature of field lines becomes significant. Therefore, the
generalization of the present approach will be the next objective of
our future work.

The next approximation concerns the fact that according to our model
magnetic field is not influenced by plasma kinematics. On the other
hand, for real astrophysical scenarios, it is obvious that plasmas
may undergo the overall configuration of the magnetic field. For
this reason, it is very important to generalize the approach
presented here and see how the collective phenomena change the
results.

\section*{Acknowledgments}
I thank Prof. A. Rogava and Dr. F. Rieger for valuable discussions.
I also thank an anonymous referee for helpful suggestions. The
research was supported by the Georgian National Science Foundation
grant GNSF/ST06/4-096.

%%%%%%%%%%

\end{document}